\documentclass[aps,prl,reprint,superscriptaddress,floatfix,nofootinbib,longbibliography]{revtex4-2}
\usepackage[T1]{fontenc}
\usepackage[utf8]{inputenc}
\usepackage{amsmath,amssymb,bm}
\usepackage{graphicx}
\usepackage{booktabs}
\usepackage{siunitx}
\usepackage{hyperref}
\hypersetup{colorlinks=true,linkcolor=blue,citecolor=blue,urlcolor=blue}
\begin{document}

\title{Proton Radiography with Isotrajectory Optics}

\author{Dmitry Varentsov}
\affiliation{GSI Helmholtzzentrum f\"ur Schwerionenforschung, Darmstadt, Germany}
\author{Maksym Miski-Oglu}
\affiliation{Helmholtz Institute Mainz, 55099 Mainz, Germany}
\author{Zs. Major}
\affiliation{GSI Helmholtzzentrum f\"ur Schwerionenforschung, Darmstadt, Germany}
\author{Paul Neumayer}
\affiliation{GSI Helmholtzzentrum f\"ur Schwerionenforschung, Darmstadt, Germany}
\author{Martin Schanz}
\affiliation{GSI Helmholtzzentrum f\"ur Schwerionenforschung, Darmstadt, Germany}
\author{John Schmidt}
\affiliation{Los Alamos National Laboratory (LANL), New Mexico, USA}

\date{26 November 2026}

\begin{abstract}

We consider a compact isotrajectory-optics scheme for proton radiography. The calculation uses a 30--60 MeV proton interval as a reference example, typical of high-energy short-pulse laser acceleration in the TNSA regime~\cite{hornung2020phelix,major2024phelix}. The idea is to use synchronized time-dependent fields such that, for one selected energy--arrival-time branch, the proton energy changes mainly the arrival time and not the transverse image position.  The reference lattice consists of four pulsed electric quadrupoles with horizontal sign pattern $+ - + -$, magnification $M_x=M_y=-3.0$ at a detector plane of $2.57\,\mathrm{m}$, a common Fourier plane at $z_F=0.987\,\mathrm{m}$, and equal outer and equal inner lens lengths. The lenses are calculated as pulsed electrostatic quadrupoles with fringe fields, through which the voltage changes while a proton passes. For a point object and a $1\,\mathrm{mrad}$ half-acceptance the closure residual of the accepted protons over 30--60 MeV is $0.25\,\mu\mathrm{m}$ ($x$) and $0.21\,\mu\mathrm{m}$ ($y$) rms in the object plane; it is set by the relativistic energy dependence. A $0.1\,\%$ voltage error or a $10\,\mathrm{ps}$ timing error of a pulse is a larger contribution. These are numerical closure tests and tolerances of the optics, not experimental resolution predictions. This version replaces the numerical results of the first one, which were obtained for an equivalent static lattice; with the real pulses that lattice is out of focus.
\end{abstract}

\maketitle

\paragraph{Isotrajectory principle.}
Laser-driven TNSA sources provide short, broadband proton bursts.  Such beams are attractive for short-exposure radiography of transient objects.  At the same time, their velocity spread makes ordinary static imaging optics strongly chromatic.  Matyshev's isotrajectory idea is to compensate this velocity dependence by synchronizing the field with the particle arrival time~\cite{matyshev1997}.

Consider first a thin electric deflector or quadrupole at a fixed distance
$s$ from a pulsed source. In the nonrelativistic limit a particle emitted
at $t=0$ reaches the element at $t=s/v$.  Its electric transverse kick scales as
\begin{equation}
    \Delta x'_E\simeq \frac{q E_\perp L}{m v^2}.
\end{equation}
To give all velocities the same transverse kick, the electric field must
scale as $E(t)\propto v^2\propto t^{-2}$.  For an electric quadrupole one
replaces $E_\perp$ by the gradient $G(t)x$, so the same
$G(t)\propto t^{-2}$ law applies.

In other words, the field is timed to the particle arrival.  Faster
particles arrive earlier and see a stronger field. Slower particles arrive
later and see a weaker field. The integrated transverse lens strength can
then be nearly the same over the intended velocity interval.

A lens has a length, and the voltage falls while a particle passes it.  For $G\propto t^{-2}$ and $t=z/v$ the normalized strength along the lens is
\begin{equation}
    k(z)=k_c\,(z_c/z)^2 ,
    \label{eq:kz}
\end{equation}
with $z_c$ the point to which the pulse is timed (here the lens center) and $k_c$ the strength there.  It does not depend on the velocity, so all trajectories coincide in the nonrelativistic limit, but it is not the strength of a static lens: the lens is stronger at its entrance than at its exit, and its integrated strength is larger by the factor $[1-(L/2z_c)^2]^{-1}$, $1.6\,\%$ for the first lens below.  A lattice matched with static lenses is therefore out of focus for the pulsed ones.

\begin{figure}[tb]
    \centering
    \includegraphics[width=0.98\columnwidth]{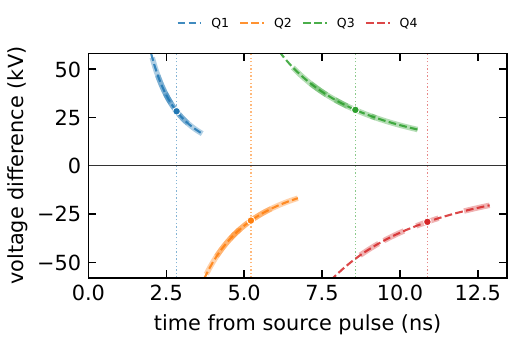}
    \caption{Lens-voltage timing for the four pulsed electric quadrupoles
    ($V$: potential difference between the $x$ and the $y$ electrodes).
    Dashed curves show the signed voltage profiles required by the relativistic
    $p v$ scaling.  Thick curves mark the voltage during the passage, from
    entrance to exit of each lens, of 30--60 MeV protons.  Dots
    mark the 40 MeV reference-proton arrival times at the lens centers.}
    \label{fig:timing}
\end{figure}

In the fully relativistic case the power law is no longer exactly $1/t^2$. The kick denominator contains $p v=\gamma m\beta^2c^2$ for electric
fields.  The corresponding ideal scaling is
$E(t),G(t)\propto\gamma\beta^2$, with $\beta=s/(ct)$ only in the simple
drift estimate. At 30--60 MeV this correction is modest, but it breaks the exact coincidence of the trajectories and sets the residual of the scheme.  The relativistic time of flight and the field along the finite lens are part of the model used here. Figure~\ref{fig:timing} shows the temporal evolution of the required voltage for the reference setup.

In the following, the 30--60 MeV interval is used as a concrete TNSA-like design case, with 40 MeV chosen as the reference energy. Here the millimetre-scale field of view is chosen as an example of compact TNSA microscopy: a small source, a small object, and a broad energy interval corrected by timing.  Larger applications require larger optics, with useful angular and energy components selected by the input aperture and Fourier-plane collimator.

The particle transport was calculated with LADA, a differential-algebra code for accelerator optics~\cite{lada2026}.  Each lens is an electrostatic quadrupole with smooth entrance and exit fringe fields whose potential satisfies Laplace's equation, including $E_z$, and whose voltage depends on time: a proton gains or loses energy in it, and the voltage changes during the passage.  Every energy has its own transfer map, third order in the coordinates (fifth order at 40 MeV changes the results by less than $1\,\%$); the closure residuals of Table~\ref{tab:closure} and the transmission are calculated from traced rays.  Independently, protons of 30, 40 and 60 MeV were traced through the same fields by direct integration of the Lorentz force; the maps and the traced rays agree within $0.04\,\mu\mathrm{m}$ at $1\,\mathrm{mrad}$ at the detector.  The field is quasistatic (see below).

Let the transverse phase-space variables be $(x,a,y,b)$, where $a=p_x/p_0$ and $b=p_y/p_0$.

\paragraph{Moment definition.}
For a given quadrupole layout, after applying the input aperture, lens-bore cuts, and Fourier collimator, the detector-plane coordinates are written as a fitted linear image plus residuals:
\begin{equation}
x_{\rm img}=x_{\rm off}+M_x x_0+r_x,
\qquad
y_{\rm img}=y_{\rm off}+M_y y_0+r_y .
\end{equation}
Here $x_{\rm off},y_{\rm off}$ are fitted image offsets, $M_x,M_y$ are fitted magnifications, and $r_x,r_y$ are the residual image errors after subtracting the best linear point-to-point image.  The residuals are evaluated particle by particle after subtracting the fitted linear image; their moments therefore include the accepted source-momentum
distribution, residual mismatch, and higher-order aberrations of the transport. All averages are taken only over particles transmitted through the stated apertures; for this the residuals are calculated from rays traced through the fields and tested at every aperture, not from the maps.

The ideal optical closure residual is then defined from the second central moment of this residual distribution, referred back to the object plane,
\begin{equation}
    \sigma_{x,\mathrm{cl}}=\frac{\sqrt{\langle r_x^2\rangle}}{|M_x|},
    \qquad
    \sigma_{y,\mathrm{cl}}=\frac{\sqrt{\langle r_y^2\rangle}}{|M_y|} .
\end{equation}
Fourth central moments are monitored separately as a tail diagnostic,
\begin{equation}
    H_x=\frac{\langle r_x^4\rangle}{\langle r_x^2\rangle^2},
    \qquad
    H_y=\frac{\langle r_y^4\rangle}{\langle r_y^2\rangle^2} .
\end{equation}
The fourth moment is kept as a separate diagnostic because the RMS closure residual alone does not describe this residual distribution. Large fourth moments indicate tails or halo that can affect contrast even when the RMS closure residual is small. This follows the statistical beam-shaping logic of Ref.~\cite{varentsov2005}, here applied to high-order maps of the final image.

The first-order transfer from the object plane to an intermediate plane $z$
is written as
\begin{align}
    x(z)&=A_x(z)x_0+B_x(z)a_0+\cdots,\\
    y(z)&=A_y(z)y_0+B_y(z)b_0+\cdots .
\end{align}
Here $A$ and $B$ denote the coordinate-to-coordinate and
transverse-momentum-to-coordinate transfer coefficients.  A common
transverse-momentum selection plane satisfies
\begin{equation}
    A_x(z_F)\simeq A_y(z_F)\simeq0,
    \qquad B_x(z_F)\simeq B_y(z_F).
\end{equation}
This is the criterion used to identify $z_F$: a finite stop there selects transverse momentum rather than object position in both transverse planes.

\paragraph{Reference lattice.}
For the reference calculation the varied quantities are the four quadrupole positions, lengths, strengths, and the Fourier-plane coordinate.  The object plane is at $z=0$, the detector plane is at $z=2.57\,\mathrm{m}$, and $r_0=3\,\mathrm{mm}$ is the bore radius.  The first-order imaging conditions are matched for the 40 MeV proton with the real pulses, Eq.~(\ref{eq:kz}) and the fringe fields included; the positions and lengths are then rounded to $0.1\,\mathrm{mm}$ and the voltages matched once more. The hard constraints are point-to-point imaging at the detector and $M_x\simeq M_y\simeq -3$, chosen as a compromise between image size and the $3\,\mathrm{mm}$ bore radius. Soft penalties reduce and equalize the quadrupole voltages, keep accepted rays inside the lens bores, place the Fourier plane after the last lens, and include both the upstream input aperture and the Fourier-plane collimator.

The selected layout has horizontal focusing signs $+ - + -$.  The two inner quadrupoles are longer than the two outer quadrupoles, with $L_1=L_4$ and $L_2=L_3$ imposed.  This choice lowers the reference voltages and makes them nearly equal. It is important to note that this symmetry is mechanical only.  Paired lenses are not dynamically identical in the pulsed case.  Different energies reach Q1--Q4 at different absolute times and spend a finite time in each lens, so the effective kick in Q1 need not match Q4, and Q2 need not match Q3.

\begin{figure*}[tb]
    \centering
    \includegraphics[width=0.94\textwidth]{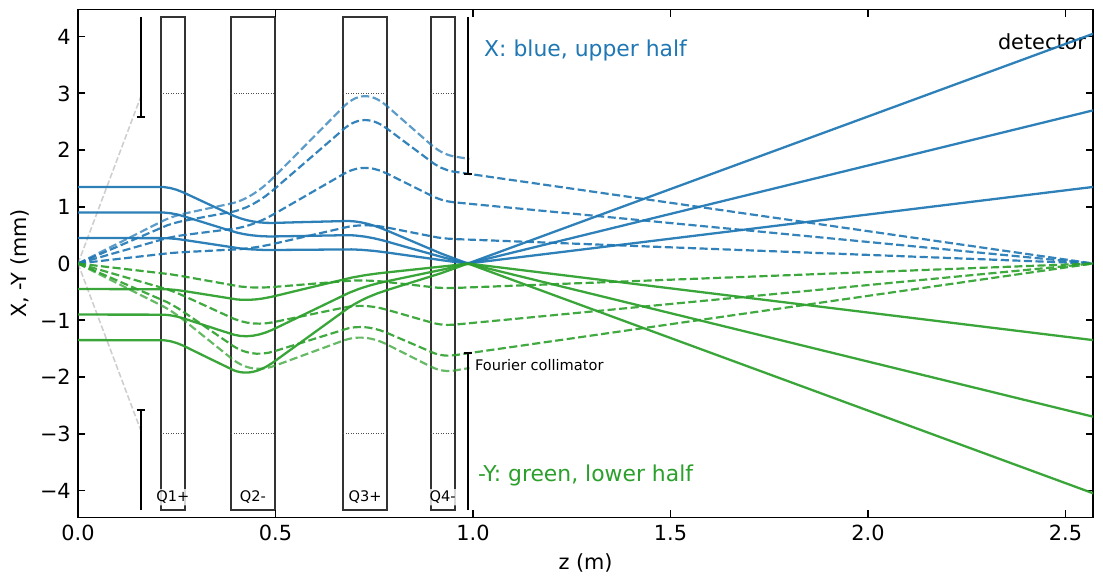}
    \caption{Reference lattice and first-order ray traces of the 40 MeV proton through the pulsed lenses, in the separated $z$--$X$ and $z$--$(-Y)$ projections.  Horizontal trajectories are drawn in blue in the upper half-plane.  Vertical trajectories are drawn as $-Y$ in green in the lower half-plane.  Solid curves have zero initial transverse momentum and finite object coordinate; dashed curves start at the source with nonzero transverse momentum.  The unlabeled upstream slit is the input aperture.  The Fourier collimator is drawn with the same slit symbol, placed at the common Fourier plane; the dashed rays with excessive source transverse momentum are stopped there.}
    \label{fig:layout}
\end{figure*}

Figure~\ref{fig:layout} shows both functions of the layout: point-to-point imaging for object-coordinate rays and transverse-momentum selection at the common Fourier plane.

\begin{table}[tb]
\caption{Four-quadrupole reference.  $z_c$ is the lens center, $L$ the lens length, $k$ the normalized strength for the 40 MeV proton at the moment it passes the center, and $V$ the potential difference between the $x$ and the $y$ electrodes (at $\pm V/2$) at that moment, $k=V/(r_0^2\,pv/e)$ with $r_0=3\,\mathrm{mm}$.  Signs refer to horizontal focusing for $k>0$.}
\label{tab:lattice}
\begin{ruledtabular}
\begin{tabular}{cccccc}
Lens & $z_c$ (m) & $L$ (mm) & $k$ ($\mathrm{m}^{-2}$) & sign & $|V|$ (kV)\\
Q1 & 0.24015 & 60.7 & +39.93 & $+$ & 28.16\\
Q2 & 0.44265 & 110.3 & -40.26 & $-$ & 28.40\\
Q3 & 0.72715 & 110.3 & +40.97 & $+$ & 28.90\\
Q4 & 0.92325 & 60.7 & -41.19 & $-$ & 29.05\\
\end{tabular}
\end{ruledtabular}
\end{table}

The resulting first-order values are
\begin{align}
    M_x &\simeq M_y \simeq -3.0, & B_x(z_D)&\simeq B_y(z_D)\simeq0,\\
    z_F &= 0.987\,\mathrm{m}, & B_x(z_F)&\simeq B_y(z_F)\simeq0.528\,\mathrm{m/rad},
\end{align}
where $z_D=2.57\,\mathrm{m}$ is the detector plane. The reference waveform convention is explicit: for lens $i$,
\begin{equation}
V_i(t)=s_i V_{i,40} \frac{p(t)v(t)}{p_{40}v_{40}} =s_i V_{i,40}
\frac{\gamma(t)\beta^2(t)}
{\gamma_{40}\beta_{40}^2},
\end{equation}
where $\beta(t)=z_{c,i}/(ct)$, $s_i=(+,-,+,-)$, $V_{i,40}$ is given in Table~\ref{tab:lattice} and $t$ is counted from the source pulse.  The 40 MeV proton passes the centers of Q1--Q4 at $t_{i,40}=(2.83,5.22,8.57,10.88)\,\mathrm{ns}$.  In the nonrelativistic drift limit this scaling reduces to the familiar $(t_{i,40}/t)^2$ law.  Between the arrivals of 60 and 30 MeV protons at the lens centers the voltage falls from about $42$--$43$ to $21$--$22\,\mathrm{kV}$.

Here $B_F\simeq B_x(z_F)\simeq B_y(z_F)$.  The Fourier-plane half-gap is
$g_F\simeq B_F\theta_c$ for half-acceptance $\theta_c$.  The input aperture
and the Fourier collimator are both transverse stops; they differ only by
position and half-gap.  The input aperture is set to a $2.58\,\mathrm{mm}$
radius at $z=0.1598\,\mathrm{m}$ so that the upstream cone does not overfill
the Q1 entrance when smooth fringe fields are included.  Table~\ref{tab:transmission}
gives candidate Fourier collimators.

For this aperture bookkeeping calculation, the source is taken as a point source with a uniform circular cone of radius $20\,\mathrm{mrad}$ in $(a,b)$. The efficiencies in Table~\ref{tab:transmission} are normalized to this selected test cone, not to the full TNSA source, and are not source-yield predictions. For a full TNSA cone with half-angle $\Theta\gg\theta_c$, the collected fraction scales approximately as $(\theta_c/\Theta)^2$ before source-yield and energy-angle correlations are included.

\begin{table}[tb]
\caption{Transmission of rays traced through the pulsed lenses and fringe fields for candidate Fourier collimators (40 MeV; with the same collimators 30 and 60 MeV differ by less than 0.02).  First-order transport agrees within 0.01. The loss columns show where particles are first removed; trans. gives the total transmission.}
\label{tab:transmission}
\begin{ruledtabular}
\begin{tabular}{cccccc}
$\theta_c$ & $g_F$ & input & bores & Fourier & trans.\\
(mrad) & (mm) & (\%) & (\%) & (\%) & (\%)\\
1 & 0.528 & 34.83 & 58.88 & 6.03 & 0.25\\
3 & 1.583 & 34.83 & 58.88 & 4.03 & 2.25\\
5 & 2.638 & 34.83 & 58.88 & 1.14 & 5.15\\
\end{tabular}
\end{ruledtabular}
\end{table}

Table~\ref{tab:closure} gives the closure residuals for a point object and the timing law above.  The source emits $a$ and $b$ Gaussian with rms $\theta_c$; the accepted part is defined by the input aperture, the bores ($|x|,|y|<r_0$ along the lenses and their fringes) and the Fourier collimator of half-gap $g_F$ of the 40 MeV protons, one stop for all energies.  It is $39$, $39$ and $33\,\%$ of the emitted protons at $\theta_c=1$, 3 and $5\,\mathrm{mrad}$, nearly the same at every energy.  At 40 MeV and $1\,\mathrm{mrad}$ the residual is $0.02\,\mu\mathrm{m}$: the fixed voltages leave a small defocus, $B_x(z_D)=+0.16$ and $B_y(z_D)=-0.09\,\mathrm{mm/rad}$ in traced rays, and the third-order aberration of the lenses adds to it in $y$ and partly cancels it in $x$.  Away from 40 MeV the relativistic part of the energy dependence defocuses: between 30 and 60 MeV $B_x(z_D)$ changes from $+1.4$ to $-2.5\,\mathrm{mm/rad}$ and $B_y(z_D)$ from $+0.9$ to $-2.2\,\mathrm{mm/rad}$, and this sets the residual of the band at $1\,\mathrm{mrad}$.  It grows linearly with $\theta_c$, the aberration with its third power; at $5\,\mathrm{mrad}$ both contribute, and the bores remove the rays with the largest aberration.  The magnification changes over the band within $\pm0.05\,\%$ ($x$) and $\pm0.15\,\%$ ($y$), that is by 0.5 and $1.5\,\mu\mathrm{m}$ at $1\,\mathrm{mm}$ from the axis.

\begin{table}[tb]
\caption{Closure residuals $\sigma_{\rm cl}$ in the object plane and tail parameters $H$ of the accepted protons: 40 MeV, and 30--60 MeV with equal emitted weights of 30, 40, 50 and 60 MeV.}
\label{tab:closure}
\begin{ruledtabular}
\begin{tabular}{ccccc}
 & \multicolumn{2}{c}{$\sigma_{x,\rm cl},\ \sigma_{y,\rm cl}$ ($\mu$m)} & \multicolumn{2}{c}{30--60 MeV}\\
$\theta_c$ (mrad) & 40 MeV & 30--60 MeV & $H_x$ & $H_y$\\
1 & 0.014, 0.020 & 0.25, 0.21 & 4.5 & 4.6\\
3 & 0.68, 0.21 & 1.2, 0.72 & 6.5 & 5.6\\
5 & 1.5, 0.95 & 2.0, 1.7 & 5.4 & 5.7\\
\end{tabular}
\end{ruledtabular}
\end{table}

\begin{table}[tb]
\caption{Image errors from errors of the four pulses and lenses, independent for each lens (quadratic sum), as $\Delta B(z_D)/|M|$: blur in the object plane per mrad of $a$ or $b$.}
\label{tab:tolerance}
\begin{ruledtabular}
\begin{tabular}{lcc}
error of each lens & $x$ ($\mu$m/mrad) & $y$ ($\mu$m/mrad)\\
voltage, $0.1\,\%$ & 3.2 & 1.6\\
pulse time, $10\,\mathrm{ps}$ & 8.1 & 5.4\\
position along the axis, $0.1\,\mathrm{mm}$ & 0.9 & 0.4\\
\end{tabular}
\end{ruledtabular}
\end{table}

Table~\ref{tab:tolerance} shows what these residuals are worth in practice.  A voltage error of $0.1\,\%$ or a timing error of $10\,\mathrm{ps}$ in each channel blurs the image by several micrometres per milliradian, more than the closure residual.  The residuals of Table~\ref{tab:closure} are limits of the scheme; the amplitude and timing stability of the pulses decides the image.

The first version of this article quoted $7.8\times10^{-4}$ and $2.6\times10^{-4}\,\mu\mathrm{m}$ and $H_x=4.31$, $H_y=4.34$ for a lattice matched with static lenses, calculated with equivalent static magnetic quadrupoles and incomplete third-order fringe terms.  With pulsed electrostatic lenses that lattice is out of focus, $B_x(z_D)=-0.069$ and $B_y(z_D)=-0.031\,\mathrm{m/rad}$ at 40 MeV, and the residual of its accepted protons at $1\,\mathrm{mrad}$ is $11.1$ and $5.1\,\mu\mathrm{m}$ (30--60 MeV).  The lattice of Table~\ref{tab:lattice} replaces it; the principle and the voltage scale are unchanged.

\paragraph{Real-field effects and validity.}
A finite electric quadrupole has fringe fields at its entrance and exit. Near the axis, a potential satisfying $\nabla^2\Phi=0$ contains longitudinal and nonlinear transverse terms, for example
\begin{equation}
    \Phi(x,y,z)=\frac{G(z)}{2}(x^2-y^2)
    -\frac{G''(z)}{24}(x^4-y^4)+\cdots ,
\end{equation}
which produces $E_z$ and higher-order transverse forces in the fringe region.  The calculation contains this series to all orders that enter the maps, with a generic fall-off of the gradient; the field of real electrodes will differ.

The fields are treated as quasistatic.  A changing voltage drives a charging current along the electrodes and with it a magnetic field; its force relative to the electric one is of order $\beta L/(c\,t)\simeq2\,\%$ at the ends of a lens fed in its middle, with opposite signs at the two ends.  A voltage step also needs up to $L/c\simeq0.4\,\mathrm{ns}$ along an electrode.  Both effects depend on the velocity and are of the size of the tolerances in Table~\ref{tab:tolerance}; they have to be included, with the feed geometry, through a time-dependent field calculation before a lattice is built.  The energy of a proton changes by up to about $10\,\mathrm{keV}$ in the four lenses at $5\,\mathrm{mrad}$, which is contained in the calculation.

A second essential limitation is longitudinal.  The pulsed isotrajectory condition corrects particles with a single relation between arrival time and energy.  Material in the object breaks this relation: a faster proton may lose energy and leave with the same final energy as a slower, weakly disturbed proton, while the two particles reach the lenses at different times.
Energy-loss straggling broadens this into a distribution in final energy, arrival time, and scattering angle.  No waveform depending only on arrival time can correct the whole distribution; off-branch particles contribute to blur, halo, or Fourier-plane rejection. Multiple Coulomb scattering can still provide useful angular-filter contrast, but the final image resolution for a real object requires coupled object-transport and optics calculations. The size of this term is object-dependent: weakly disturbing thin objects may remain close to the selected branch, whereas thick or high-$Z$ objects can make object-induced timing and energy spread the dominant resolution term. Thus the sub-micrometre closure values above should not be interpreted as object-resolution predictions after material interaction.

\paragraph{Pulsed-HV implementation.}
The required driver is a synchronized nanosecond pulsed-HV system.  Each channel drives one small capacitive quadrupole, and the relevant optics quantity is the measured voltage history at the electrode gap.  In the present example each channel has to cover roughly 21--43 kV over a few nanoseconds with the polarity given in Table~\ref{tab:lattice}.

A simple current estimate sets the driver scale.  For a capacitance $C$ and a voltage change $\Delta V$ over a time $\Delta t$, the capacitive current is $I_C\simeq C\Delta V/\Delta t$.  For a $1$--$10\,\mathrm{pF}$ electrode system and tens of kilovolts over a few nanoseconds this gives transient currents from order $10\,\mathrm{A}$ to a few $100\,\mathrm{A}$.  This current scale is demanding but feasible for a dedicated pulsed-power driver; it makes impedance matching, reflections, feedthrough response, ringing, jitter, polarity, repetition rate, and direct gap-voltage measurement part of the design.

These issues are not separate from the optics design.  Aperture, lens length, energy interval, and voltage waveform all trade against each other.  If the voltage scale is too aggressive, the natural mitigations are smaller aperture, longer lenses, a narrower energy interval, or a readjusted lattice.

\paragraph{Practical interpretation.}
The numbers above are an imaging-optics feasibility test under idealized conditions.  They show that a compact four-lens system with pulsed electrostatic lenses can form a point-to-point image for a selected time--energy component and with a common Fourier plane, provided that amplitude and timing of the pulses are controlled to the level of Table~\ref{tab:tolerance}.  The voltage scale quoted here belongs to the present model and to the stated $3\,\mathrm{mm}$ bore radius.  The calculation does not include a measured TNSA source, target-dependent scattering and energy loss, detector response, waveform distortions, electrode manufacturing tolerances, or space-charge fields.  These effects are experiment-specific and have to be added after a concrete object, source spectrum, detector concept, and pulsed-HV prototype have been chosen.

The calculation therefore has a precise scope.  Pulsed isotrajectory optics can reduce chromatic image blur for a selected time--energy component of a broadband proton burst.  It is not a universal correction for all particles produced by a laser target or by subsequent object interaction. For thin objects in a TNSA-microscopy geometry, the selected time--energy component can be imaged with Fourier-plane angular filtering; for thick degraders, object-induced energy loss and straggling can dominate the blur and must be treated in a separate object-transport calculation.  A quantitative comparison with static matched optics is therefore object-dependent and requires the same source, aperture, and material model.

The next useful step is a coupled study of source model, object physics, pulser response, and electrode fields, rather than further optimization of the ideal map alone.  For larger-field applications the same principle requires larger bore radii, longer lenses, and a different magnification.  A natural extension is to study isotrajectory optics as a time-dependent imaging module for ELIMAIA/ELIMED, where energy selection, dosimetry, and sample irradiation are already part of the beamline concept~\cite{cirrone2020elimed}.

\begin{acknowledgments}
This working calculation is part of a proton-radiography optics study at GSI.
\end{acknowledgments}

\end{document}